\documentclass[preprint,12pt]{elsarticle}

\usepackage{graphicx}
\usepackage{amssymb}
\usepackage{amsmath}
\usepackage{changes}
\newcommand{\imp}{\text{imp}}
\newcommand{\hyb}{\text{hyb}}
\newcommand{\con}{\text{con}}
\newcommand{\bra}[1]{\langle #1 |}
\newcommand{\ket}[1]{| #1 \rangle}

\newcounter{bla}

\journal{Computer Physics Communications}

\begin{document}

\begin{frontmatter}



\title{Numerical renormalization group calculations for
magnetic impurity systems with spin-orbit coupling and
crystal-field effects}


\author[a,b]{Aitor Calvo-Fernández\corref{author}}
\author[d,b,e]{María Blanco-Rey}
\author[a,b,c]{Asier Eiguren}

\cortext[author] {Corresponding author.\\\textit{E-mail
address:} aitor.calvof@ehu.eus}
\address[a]{Departamento de Física, Universidad del País 
Vasco UPV/EHU, 48080 Leioa, Spain}
\address[b]{Donostia International Physics Center (DIPC),
20018 Donostia-San Sebastián, Spain}
\address[c]{EHU Quantum Center, Universidad del País Vasco
UPV/EHU, 48080 Leioa, Spain}
\address[d]{Departamento de Polímeros y Materiales
Avanzados: Física, Química y Tecnología, Universidad del
País Vasco UPV/EHU, 20018 Donostia-San Sebastián, Spain}
\address[e]{Centro de Física de Materiales CFM/MPC
(CSIC-UPV/EHU), 20018 Donostia-San Sebastián, Spain}

\begin{abstract}
Exploiting symmetries in the numerical renormalization group
(NRG) method significantly enhances performance by improving
accuracy, increasing computational speed, and optimizing
memory efficiency. Published codes focus
on continuous rotations and unitary groups, which generally are
not applicable to systems with strong
crystal-field effects. The \verb|PointGroupNRG| code
implements symmetries related to discrete rotation groups,
which are defined by the user in
terms of Clebsch-Gordan coefficients, together with particle
conservation and spin rotation symmetries. In this paper we
present a new version of the code that extends the available
finite groups, previously limited to simply reducible point
groups, in a way that all point and double groups become
accessible. It also includes the
full spin-orbital rotation group. Moreover, to improve the code's
flexibility for impurities with complex interactions, this
new version allows to choose between a standard Anderson
Hamiltonian for the impurity or, as another novel feature, an
ionic model that requires only the spectrum and the impurity
Lehmann amplitudes.

% A submitted program is expected to satisfy the following
% criteria: it must be of benefit to other physicists, or be
% an exemplar of good programming practice, or illustrate new
% or novel programming techniques which are of importance to
% computational physics community; it should be implemented in
% a language and executable on hardware that is widely
% available and well documented; it should meet accepted
% standards for scientific programming; it should be
% adequately documented and, where appropriate, supplied with
% a separate User Manual, which together with the manuscript
% should make clear the structure, functionality,
% installation, and operation of the program.
%
% Your manuscript and figure sources should be submitted through Editorial Manager (EM) by using the online submission tool at \\
% https://www.editorialmanager.com/comphy/.
%
% In addition to the manuscript you must supply: the program source code; a README file giving the names and a brief description of the files/directory structure that make up the package and clear instructions on the installation and execution of the program; sample input and output data for at least one comprehensive test run; and, where appropriate, a user manual.
%
% A compressed archive program file or files, containing these items, should be uploaded at the "Attach Files" stage of the EM submission.
%
% For files larger than 1Gb, if difficulties are encountered during upload the author should contact the Technical Editor at cpc.mendeley@gmail.com.

\end{abstract}


\end{frontmatter}

% \linenumbers

% All CPiP articles must contain the following
% PROGRAM SUMMARY.

{\bf PROGRAM SUMMARY}
  %Delete as appropriate.

\begin{small}
\noindent
{\em Program Title:} PointGroupNRG                                          \\
{\em CPC Library link to program files:} (to be added by Technical Editor) \\
{\em Developer's repository link:} https://github.com/aitorcf/PointGroupNRG \\
{\em Code Ocean capsule:} (to be added by Technical Editor) \\
{\em Licensing provisions:} GPLv3 \\
% {\em Licensing provisions(please choose one) :} CC0 1.0/CC By 4.0/MIT/Apache-2.0/BSD 3-clause/BSD 2-clause/GPLv3/GPLv2/LGPL/CC BY NC 3.0/MPL-2.0  \\
{\em Programming language:} Julia                                   \\ 
% {\em Supplementary material:}                                 \\
  % Fill in if necessary, otherwise leave out.
{\em Journal reference of previous version:} https://doi.org/10.1016/j.cpc.2023.109032\\
{\em Does the new version supersede the previous version?:}
Yes\\
  %Only required for a New Version summary, otherwise leave out.
{\em Reasons for the new version:} Extension\\
  %Only required for a New Version summary, otherwise leave out.
%{\em Summary of revisions:}*\\
  %Only required for a New Version summary, otherwise leave out.
% (approx. 50-250 words)
{\em Nature of problem:}
Numerical renormalization group (NRG) calculations for
realistic models are computationally expensive, mainly due
to their hard scaling with the number of orbital and spin
configurations available for the electrons. Symmetry
considerations reduce the computational cost of the
calculations by exploiting the block structure of the
operator matrix elements and by removing the redundancy in
the symmetry-related matrix elements. Existing codes
implement continuous symmetries, which are not generally
and/or straightforwardly applicable to systems where
spin-orbit and crystal-field effects need to be taken into
account.\\
  %Describe the nature of the problem here. \\
% (approx. 50-250 words)
{\em Solution method:} % (approx. 50-250 words)
The first version of the code \cite{calvo-fernandez2024}
introduced finite point group symmetries together with
particle conservation and spin isotropy, useful for systems
with strong crystal-field effects but negligible spin-orbit
coupling. This new version also includes total angular
momentum conservation and double group symmetries, together
with particle conservation. This allows to deal with magnetic
impurity systems with strong spin-orbit coupling. To make
the code more versatile in handling systems of this type, we
have added the possibility to use ionic models.\\
  %Describe the method solution here.
% {\em Additional comments including restrictions and unusual features (approx. 50-250 words):} \\
%
%    \\

% \begin{thebibliography}{0}
% \bibitem{1}Reference 1         % This list should only contain those items referenced in the                 
% \bibitem{2}Reference 2         % Program Summary section.   
% \bibitem{3}Reference 3         % Type references in text as [1], [2], etc.
%                                % This list is different from the bibliography at the end of 
%                                % the Long Write-Up.
% \end{thebibliography}
\end{small}

\section{Introduction}
The numerical renormalization group (NRG) is a
non-perturbative and essentially exact iterative procedure
designed to accurately describe the low-temperature physics
of magnetic impurity models \cite{bulla2008}. One of its
main limitations is the exponential increase of the
computational effort as the number of orbital and spin
configurations of the electrons increase. Exploiting the
symmetry of the models is a common strategy that allows to
reduce this scaling and thereby render complex systems more
tractable. Historically, continuous unitary and rotation
symmetries have already been included in the literature
since the first implementations for the Anderson model
\cite{krishna-murthy1980} and have subsequently been
developed \cite{weichselbaum2012} and incorporated in
publicly available codes \cite{nrgljubljana, zitko2009,
toth2008}.
\par
In a previous work \cite{calvo-fernandez2024} we presented
\texttt{PointGroupNRG}, a NRG code written in Julia
language. Its main feature was the implementation of simply
reducible finite point group symmetries together with charge
and spin symmetry, which allows to construct and efficiently
solve models for systems including crystal-field effects.
Here we introduce a new version of \texttt{PointGroupNRG}
with an extended functionality allowing to deal also with
systems where spin-orbit effects are important. This
involves two additional types of symmetry, the spin-orbital
$SU(2)$ and the double groups, for systems where the
spin-orbit coupling is the sole symmetry-breaking effect or
it is accompanied by a crystal field. In an effort to
deal with the relevant double group symmetries and in order
to make the code more widely applicable, this new version
works with any point and double group, overcoming the
limitation of the previous version, which was applicable
only to simply reducible groups. Now the code also
implements ionic models, which are suited for impurities
where correlations and symmetry-breaking terms result in a
complicated Hamiltonian that is better dealt with by
retaining only the lowest-lying multiplets. With these new
features, the \verb|PointGroupNRG| code can now be used to
study systems with interesting Kondo physics arising from
spin-orbit coupling and crystal-field effects, such as
heavy-fermion materials \cite{cox1987,bib:koga99,hotta2018}
or low-dimensional $d$-shell impurity systems
\cite{roura-bas2014,dinapoli2013a,dinapoli2013b,dinapoli2015},
by means of a NRG procedure adapted to their double group
symmetry.
\par
The paper is organized as follows. Section
\ref{theoretical_background} is devoted to the theoretical
elements on which the new functionality of the code is
based, including the ionic model (\ref{ionic_model}) and the
elements of symmetry
(\ref{symmetry_breaking_scheme}-\ref{beyond_simply_reducible_groups}).
In Section \ref{implementation} we describe the main novel
features of the code interface
(\ref{clebschgordan_coefficients} and
\ref{model_definition}), also pointing out some key
modifications that have been introduced to improve the
iterative procedure (\ref{iterative_procedure}). We finish
with Section \ref{example}, where we work out the example of
a uranium impurity model by showing the usage (\ref{usage}),
followed by entropy calculation results and performance
benchmarks (\ref{results_and_performance})

\section{Theoretical
background}\label{theoretical_background}

\subsection{Ionic model}\label{ionic_model}
The Anderson Hamiltonian $\mathcal H$ for magnetic impurity
systems is the sum of the impurity term, the conduction
electron term, and the hybridization,
\begin{equation}
\begin{aligned}
    \mathcal H
    =&
    H_\imp
    +
    H_\con
    +
    H_\hyb.
\label{eq:anderson_hamiltonian}
\end{aligned}
\end{equation}
We write the impurity part in the diagonal basis as
\begin{equation}
    H_\imp
    =
    \sum_{i} \epsilon_i \ket{i}\bra{i},
\end{equation}
where $\epsilon_i$ are the eigenenergies and $\ket{i}$ the
eigenstates of the free impurity. The band of conduction
electrons is described by
\begin{equation}
    H_\con
    =
    \sum_{b} 
    \int_{-D}^D
    \epsilon
    c^\dagger_{\epsilon b}
    c_{\epsilon b}
    {\rm d}\epsilon,
\end{equation}
where $b$ is a set of quantum numbers for the electronic
states, $D$ is the half-bandwidth, and $c^\dagger_{\epsilon
b}$ ($c_{\epsilon b}$) is the creation (annihilation)
operator for an electron with energy $\epsilon$ and quantum
numbers $b$. The impurity and the conduction electrons are
coupled by the hybridization term
\begin{equation}
    H_\hyb
    =
    \sum_{i,j}\sum_{ a, b}\int_{-D}^D
    {\rm d}\epsilon
    [\rho_{ b}(\epsilon)]^\frac{1}{2}V_{ a b}
    \bra{i} f^\dagger_{ a} \ket{j}
    \ket{i}\bra{j}
    c_{\epsilon b}
    + 
    \text{h.c.},
\label{eq:hybridization}
\end{equation}
where $\rho_{ b}(\epsilon)$ is the density of states in the
channel with quantum numbers $b$, $V_{a b}$ is the tunneling
amplitude between impurity and conduction states with
quantum numbers $a$ and $b$, respectively, and $f_ a$
is the annihilation operator for an impurity electron with quantum numbers $a$.
The electronic structure of the impurity enters the
Hamiltonian through the Lehmann amplitudes
$\bra{i}f^\dagger_{a}\ket{j}$. This form of the Anderson
Hamiltonian, with $H_\text{imp}$ and $H_\text{hyb}$ written
in terms of the free impurity eigenstates $\ket{i}$, is
known as the ionic model and is appropriate for cases with
complex impurity configurations with strong spin-orbit
coupling and/or crystal-field effects
\cite{hirst1978,hewson1997}.
\par

\subsection{Symmetry in the
Hamiltonian}\label{symmetry_in_the_hamiltonian}
To make the Hamiltonian reflect the symmetry of the system,
which is given by a symmetry group $G$, we work in a
symmetry-adapted basis where the sets of quantum numbers
are decomposed as
\begin{equation}
\begin{aligned}
    u := (\Gamma_u,\gamma_u,r_u),
\end{aligned}
\end{equation}
where $\Gamma_u$ is the irreducible representation (irrep),
$\gamma_u$ is the partner number, and $r_u$ is an outer
multiplicity label distinguishing states with the same
symmetry. Then $\mathcal H$ can be made totally symmetric
under $G$ by choosing $\epsilon_i$,
$\rho_{b}(\epsilon)$ and $V_{ab}$ to have
the form,
\begin{align}
    &\epsilon_i = \epsilon(\Gamma_i)_{r_i},
    \label{eq:energies}
    \\
    &\rho_b (\epsilon) =
    \rho(\Gamma_b;\epsilon)_{r_b}
    \label{eq:dos}
    \\
    &V_{ab} = 
    \delta_{\Gamma_a \Gamma_b}
    \delta_{\gamma_a \gamma_b}
    V(\Gamma_a)_{r_a r_b}.
    \label{eq:tunneling}
\end{align}
This ensures that (i) states belonging to the same multiplet
are degenerate and (ii) states labeled by different quantum
numbers, which are assumed to form an orthogonal basis, only
mix if they have the same irrep and partner quantum numbers
as required by symmetry \cite{hamermesh1989}.

\subsection{Symmetry-breaking
scheme}\label{symmetry_breaking_scheme}
In the absence of any symmetry-breaking effects, we take
$\mathcal H$ to be invariant under the action of operations
in the symmetry group
\begin{equation}
    G_\text{UOS} = 
    U(1)_\text{C} \otimes 
    SO(3)_\text{O} \otimes
    SU(2)_\text{S},
\label{eq:G_UOS}
\end{equation}
which is the outer direct product of $U(1)$ charge symmetry
(particle conservation), $SO(3)$ orbital symmetry
(conservation of total orbital angular momentum), and
$SU(2)$ spin symmetry (conservation of total spin). This
symmetry is typically broken to some degree in real systems
such as $d$-shell impurities in metallic environments where
the spin-orbit coupling can be neglected, in which case
crystal field effects reduce the orbital symmetry from
$SO(3)_\text{O}$ to the point group $P_\text{O}$, resulting
in a symmetry group 
\begin{equation}
G_\text{UPS}=U(1)_\text{C} \otimes
P_\text{O} \otimes SU(2)_\text{S}.
\end{equation}
This case was worked out extensively in the previous paper
\cite{calvo-fernandez2024}, where we also covered the
construction of $H_\imp$ starting from the on-site energies
and Coulomb repulsion terms in the symmetry-adapted scheme. 
\par
If the spin-orbit coupling is taken into account, the
symmetry is reduced from $G_\text{UOS}$ to
\begin{equation}
    G_\text{UJ} = U(1)_\text{C} \otimes SU(2)_\text{OS},
\end{equation}
which expresses invariance under simultaneous rotations in
the orbital and spin coordinates, which belong to the group
$SU(2)_\text{OS}$. The appropriate symmetry quantum numbers
for $G_\text{UJ}$ are $N$ (particle number), $J$ (total
angular momentum) and $j$ (total angular momentum
projection). If we add crystal-field effects in addition to
the spin-orbit coupling, we end up with the symmetry
\begin{equation}
    G_\text{UD} = U(1)_\text{C}\otimes D_\text{OS},
\end{equation}
where $D_\text{OS}$ is a double group representing discrete
rotations acting simultaneously on the orbital and spin
spaces. The appropriate symmetry quantum numbers in this
case are $N$, $I$ ($D_\text{OS}$ irrep) and $i$
($D_\text{OS}$ partner number). This description of the
symmetry in the presence of spin-orbit coupling applies
without loss of generality to the \textit{L-S}, intermediate
and $j-j$ coupling regimes \cite{bib:moore09}.

\subsection{Beyond simply reducible groups}
\label{beyond_simply_reducible_groups}
The unitary group $U(1)$, the full rotation groups $SO(3)$,
$O(3)$ and $SU(2)$, and some relevant point groups, for
instance the cubic group $O$, are simply reducible. This
means that the in the decomposition of the inner direct
product of any two irreps into a direct
sum of irreps \cite{bradley2010},
\begin{equation}
    \Gamma_i\boxtimes\Gamma_j
    =
    \oplus_{\Gamma_k} n_k \Gamma_k,
\label{eq:irrep_decomposition}
\end{equation}
each irrep $\Gamma_k$ appears at most only once and
therefore $n_k=1$ or 0 for all $k$ \cite{hamermesh1989}. When
considering double groups, however, we find that in most of
the relevant cases this condition is not satisfied, such as for
the double group $\tilde O$ that is the double cover of $O$
\cite{altmann1994}.
\par
For our purposes, the direct consequence of going beyond
simply reducible groups is that the Clebsch-Gordan
coefficients carry an additional index that labels the
distinct subspaces belonging to the same irrep arising from
the decomposition of the product of states or operators.
This applies to the product of multiplet states,
\begin{equation}
    \ket{i}
    \otimes
    \ket{j}
    =
    \sum_{\Gamma_k,\gamma_k}\sum_{\alpha_k}
    (\Gamma_i,\gamma_i;\Gamma_j,\gamma_j|\Gamma_k,\gamma_k)_{\alpha_k}^*
    \ket{k;\alpha_k},
\label{eq:clebschgordan_states}
\end{equation}
where $(\Gamma_i,\gamma_i; \Gamma_j,\gamma_j|
\Gamma_k,\gamma_k) _{\alpha_k}$ is a Clebsch-Gordan
coefficients with the additional index $\alpha_k$, following
the notation in Ref. \cite{moca2012}. This implies some
freedom in defining the Clebsch-Gordan coefficients and the
states in the decomposition for $\Gamma_k$ with $n_k>1$,
depending on the choice of the $\alpha_k$ subspaces.
Although the quantum numbers $k$ fully identify the new
states, the index $\alpha_k$ provides additional information
that we explicitly include in the notation whenever
necessary. 
\par 
The additional index also applies to the Wigner-Eckart theorem:
using the matrix elements of the impurity electron operators
as an example and adopting the compact notation for the
multiplets $m_u=(\Gamma_u,r_u)$, we have
\begin{equation}
    \bra{i} f^\dagger_a \ket{j}
    =
    \sum_{\alpha}
    \bra{m_i}| 
    f^\dagger_{m_a} 
    |\ket{m_j}_\alpha
    (\Gamma_a,\gamma_a;
    \Gamma_j,\gamma_j|
    \Gamma_i,\gamma_i)^*_\alpha,
\label{eq:wigner-eckart}
\end{equation}
where $\bra{m_i}| f^\dagger_{m_a} | \ket{m_j}_\alpha$ is a
reduced matrix element. The index $\alpha$ here
distinguished the subspaces belonging to $\Gamma_i$ in the
decomposition of $\Gamma_a \boxtimes \Gamma_j$. Conversely,
the reduced matrix elements are obtained by inverting Eq.
\ref{eq:wigner-eckart} using the orthogonality of the
Clebsch-Gordan coefficients \cite{hamermesh1989},
\begin{equation}
    \bra{m_i}|f_{m_a}|\ket{m_j}_\alpha
    =
    \sum_{\gamma_a,\gamma_i}
    \bra{i}f_{a}\ket{j}
    (\Gamma_a,\gamma_a;\Gamma_i,\gamma_i|\Gamma_j,\gamma_j)_\alpha^*.
\label{eq:wigner-eckart_inverted}
\end{equation}

\section{Implementation}\label{implementation}

\subsection{Clebsch-Gordan coefficients}
\label{clebschgordan_coefficients}
\verb|PointGroupNRG| can handle any symmetry of the
structure $G_\text{UPS}$, $G_\text{UJ}$ or $G_\text{UD}$.
The only requirements from the user are the Clebsch-Gordan
coefficients for the finite point or double groups, as the
Clebsch-Gordan coefficients for the continuous groups are
calculated by the program. Sanity checks are performed
automatically on the coefficients provided by the user to
inform of possible errors. 
\par
The input format of the coefficients is user-friendly to
facilitate their manual entry and correction. For a group
\texttt{G}, a directory \texttt{GCG} (any name is valid)
containing the Clebsch-Gordan coefficients is required. The
user is assumed to provide coefficients corresponding to the
decomposition of the product of a pair of irreps in text
files named \texttt{NxM\_AxB.txt}. \texttt{A} and \texttt{B}
are the user-defined strings for the \texttt{N}-th and
\texttt{M}-th irreps of \texttt{G}, respectively, starting
from 0. Each paragraph in those files should contain the
information about one of the irreps appearing in the
decomposition of \texttt{A}$\boxtimes$\texttt{B}, following
the format
\begin{verbatim}
    L C a
    ( 1 1 | 1 ) = <(A, 1; B, 1 | C, 1)>
    ...
    ( n m | l ) = <(A, n; B, m | C, l)>
\end{verbatim}
where the paragraph corresponds to the \texttt{a}-th
subspace of the \texttt{L}-th irrep \texttt{C} of
\texttt{G}. \texttt{n}, \texttt{m} and \texttt{l} are the
dimensions of \texttt{A}, \texttt{B} and \texttt{C},
respectively. The segment \texttt{<(A, 1; B, 1 | C, 1)>} has
to be substituted by the corresponding Clebsch-Gordan
coefficient given in any format that can be parsed in Julia
as a complex number. It suffices to provide only the non-zero
coefficients. As an example, the code includes the
Clebsch-Gordan coefficients for the double group $\tilde O$.

\subsection{Model definition}
\label{model_definition}
\texttt{PointGroupNRG} provides a flexible interface for
constructing custom models to be solved with the NRG. This
is achieved by specifying a symmetry group for the system
and then constructing $H_\text{imp}$, $H_\text{con}$ and
$H_\text{hyb}$ separately. The three terms are defined in a
symmetry-adapted basis so that the code can use them to
generate a completely symmetric Hamilonian.
\par
The impurity term can be defined in two ways. The first is
to provide the available electronic states, the on-site
energies and the interaction between the electrons, which
are used to construct the appropriate basis and the
Hamiltonian. This procedure is described in the previous
publication \cite{calvo-fernandez2024}. The second option,
which is a new feature in this version, is to define the
ionic model as described in Section \ref{ionic_model} by
providing the impurity spectrum and the tunneling
amplitudes, using the dictionary format of Julia language. A
detailed description of the input needed is provided in
Section \ref{example}.
\par
The definition of the remaining Hamiltonian terms remains the
same as in the first version, with a format similar to
the impurity part. The user must provide the density of states
functions and the tunneling amplitudes in Eqs. \ref{eq:dos}
and \ref{eq:tunneling}, respectively. Additionally, the
conduction electron degrees of freedom need to be specified
and a preparatory calculation has to be run in order to
compute the conduction shell multiplet states.

\subsection{Iterative procedure}
\label{iterative_procedure}
Symmetry is exploited in the NRG calculation by representing
every operator by its reduced matrix elements, which allows
for faster and more memory-efficient calculations
\cite{calvo-fernandez2024}. The remaining information is
stored in the form of irrep decompositions and sums over
Clebsch-Gordan coefficients, which allows to construct the
multiplet basis at each step in the iteration, and the
required operators in that basis. 
\par
In the following we give the main expressions, which are
similar to those used in the previous version
\cite{calvo-fernandez2024} but now they include the indices
$\alpha_u$ (also $\alpha_v$) from Eq. \ref{eq:clebschgordan_states} and
$\alpha$ (also $\beta$) from Eqs. \ref{eq:wigner-eckart} and
\ref{eq:wigner-eckart_inverted} in order to accommodate
also the groups that are not simply reducible.
\par
At the $n$-th step of the NRG procedure, we first construct
the basis of multiplets $m_u$ ($m_v$) by combining the
multiplets $m_i$ ($m_j$) from the diagonal basis of the
previous step and $m_\mu$ ($m_\nu$) from the $n$-th
conduction electron shell. Taking their tensor products, we
obtain the new basis as 
\begin{equation}
    m_i \otimes m_\mu = \oplus_{m_u} (m_u)_{\alpha_u},
\end{equation}
where each irrep $\Gamma_u$ will be repeated $n_u$ times in
the multiplets $m_u$ according to Eq.
\ref{eq:irrep_decomposition}, each with a different
$\alpha_u$ as in Eq. \ref{eq:clebschgordan_states}.
\par
The reduced matrix elements for the operators are
constructed using the multiplets $m_u$ and sums over
Clebsch-Gordan coefficients. For the hopping operator
$c^\dagger_{n-1,a} c_{n,a}$ that couples the already
included degrees of freedom to the $n$-th shell, we have
\begin{equation}
\begin{aligned}
    \bra{m_u;\alpha_u}|
    c^\dagger_{n-1,m_a}c_{n,m_a}
    |\ket{m_v;\alpha_v}
    =
    \delta_{\Gamma_u \Gamma_v}
    \sum_{m_a} \sum_{\alpha,\beta}
    \bra{m_i}|
    c^\dagger_{n-1,m_a}
    |\ket{m_j}_\alpha^{[n-1]}
    & \\ \times
    (\bra{m_\nu}|
    c^\dagger_{n,m_a}
    |\ket{m_\mu}_\beta^{[n]})^*
    D(\Gamma_u,\Gamma_a,\Gamma_i,\Gamma_j,\Gamma_\mu,\Gamma_\nu)
    _{\alpha_u,\alpha_v,\alpha,\beta} &.
\label{eq:D}
\end{aligned}
\end{equation}
For the $n$-th shell electronic operator $c^\dagger_{n,m_a}$,
the expression is
\begin{equation}
\begin{aligned}
    \bra{m_u;\alpha_u}|
    c^\dagger_{n,m_a}
    |\ket{m_v;\alpha_v}_\beta ^{[n]}
    =
    \delta_{m_i m_j}
    \sum_{\alpha}
    \bra{m_\mu}|
    c^\dagger_{n,m_a}
    |\ket{m_\nu}_\alpha &
    \\ \times
    K(\Gamma_u,\Gamma_v,\Gamma_a,\Gamma_i,\Gamma_\mu,\Gamma_\nu)
    _{\alpha_u,\alpha_v,\alpha,\beta}. &
\label{eq:K}
\end{aligned}
\end{equation}
Lastly, the impurity operator $f^\dagger_{m_a}$ is computed
as
\begin{equation}
\begin{aligned}
    \bra{m_u;\alpha_u}|
    f^\dagger_{m_a}
    |\ket{m_v;\alpha_v}^{[n]} _{\beta}
    =
    \delta_{m_\mu m_\nu}
    \sum_{\alpha}
    \bra{m_i}|
    f^\dagger_{m_a}
    |\ket{m_j}_\alpha &
    \\ \times
    F(\Gamma_u,\Gamma_v,\Gamma_a,\Gamma_i,\Gamma_j,\Gamma_\mu)
    _{\alpha_u,\alpha_v,\alpha,\beta}. &
\label{eq:F}
\end{aligned}
\end{equation}
The Clebsch-Gordan coefficient sums $D(\dots)_{\dots}$,
$K(\dots)_{\dots}$ and $F(\dots)_{\dots}$ are computed
according to the formulas in Ref. \cite{moca2012}, where the
symmetry-adapted iterative procedure was
explained in detail. Within the \verb|PointGroupNRG| code, the
Clebsch-Gordan sums are computed at the beginning of the
calculation and stored in sparse \texttt{Dict} containers,
which greatly reduces the computational cost.

\section{Example: quadrupolar Kondo effect}\label{example}
Heavy-fermion materials featuring $f$ electrons have been
proposed as physical realizations of multichannel Kondo
systems displaying non-Fermi liquid behavior at low
temperatures. In these systems, the effect of the crystal
field on the $f$ orbital configuration is determinant. D. L.
Cox \cite{cox1987} proposed a model for UBe$_{13}$
consisting
of a two-channel Kondo Hamiltonian with exchange
interactions on the quadrupolar moments. The U($5f^2$) atoms
sit at sites of cubic symmetry and the spin-orbit coupling is
strong, resulting in a cubic spin-orbital symmetry $\tilde
O_\text{OS}$~\cite{bib:lea62}. In the so-called "3-7-8"
model, a non-Kramers doublet ground multiplet of $E$
symmetry (${ \ket{\Gamma_3,+}, \ket{\Gamma_3,-} }$
levels~\cite{bib:lea62}) couples to an excited multiplet of
$E_{1/2}$ symmetry (${ \ket{\Gamma_7,\pm} }$) via lectronic
excitations of a conduction electron quartet of $F_{3/2}$
symmetry (${ \ket{\Gamma_8,\pm 2}, \ket{\Gamma_8,\pm 1} }$).
This minimal model reproduces the non-Fermi liquid ground
state of UBe$_{13}$. The phase diagram of $f^2$ impurities
is shown to have another non-Fermi liquid fixed point when,
instead of the doublet, a triplet ground state is
considered, which can be accessed in a different crystal
field~\cite{bib:koga99}. Nevertheless in the following, we
choose the "3-7-8" model as the platform to showcase the
usage of \texttt{PointGroupNRG}, as it provides a simple,
yet physically relevant, scenario of an ionic model of
double group symmetry.

\subsection{Usage}\label{usage}
The first step consists on constructing the
multiplet states for the conduction shell degrees of
freedom. We achieve this by running the
\texttt{compute\_multiplets} function provided by the
\texttt{PointGroupNRG.MultipletCalculator} submodule,
\begin{verbatim}
    using PointGroupNRG.MultipletCalculator
    symmetry = "D"
    irrep = "F32"
    clebschgordan_path = "clebschgordan"
    compute_multiplets(
        symmetry;
        irrep=irrep,
        clebschgordan_path="clebschgordan"
    )
\end{verbatim}
In this example, we are indicating the
\texttt{compute\_multiplets} function to use the symmetry
type $G_\text{UOS}$ by setting \texttt{symmetry="D"} (D
stands for double group) and to compute the multiplet states
of the irrep $F_{3/2}$, denoted \texttt{F32} in the code, by
using the Clebsch-Gordan coefficients contained in the
directory \texttt{clebschgordan} in the format specified in
Section \ref{clebschgordan_coefficients}.
\par
The actual NRG calculation is executed by the
\texttt{nrgfull} function from the submodule
\texttt{PointGroupNRG.NRGCalculator}. We begin by defining
the impurity model,
\begin{verbatim}
    using PointGroupNRG.NRGCalculator
    label = "U378"
    ground  = (2,"E",0.0)
    excited = (1,"E12",0.0)
    tunnel  = (1,"F32",0.0)
    spectrum = Dict(
        ground  => [0.0],
        excited => [0.1]
    )
    lehmann_iaj = Dict(
        (ground,tunnel,excited)=>ones(ComplexF64,1,1,1,1)
    )
\end{verbatim}
With \texttt{label="U378"} we set a custom name for the model. 
The variables \texttt{ground}, \texttt{excited} and
\texttt{tunnel} represent the irreps $\Gamma_g$, $\Gamma_e$
and $\Gamma_t$ of the ground multiplet, the excited
multiplet and the tunneling electrons, respectively, as
3-tuples carrying the $U(1)_\text{C}$, $\tilde O_\text{OS}$ and
$SU(2)_\text{S}$ symmetry quantum numbers. To disable the
latter in calculations with double group symmetry, we set
the total spin quantum number to 0. $H_\imp$ is defined by
\texttt{spectrum}, which contains the energies
$\epsilon(\Gamma_i)_{r_i}$ of the multiplets belonging to
each impurity irrep. All energy-related magnitudes are given
in units of the half-bandwidth $D$. The reduced matrix
elements of the creation operator or reduced Lehmann
amplitudes are given by \texttt{lehmann\_iaj}, which has a
single entry for $\alpha=r_i=r_a=r_j=1$. By doing this, we have arbitrarily
set it to $1$ so that we can control the entire coupling
strength $V(\Gamma_t)_{11}
\bra{\Gamma_g,1}|f^\dagger_{\Gamma_t,1}|\ket{\Gamma_e,1}$
with the tunneling amplitude $V(\Gamma_t)_{11}$. 
\par
Next, we define the degenerate conduction bands of $F_{3/2}$
symmetry with $\rho(\Gamma_t;\epsilon)_1=1/2$ in units of
the half-bandwidth,
\begin{verbatim}
    shell_config = Dict("F32" => 1)
    channels_dos = Dict("F32" => [x->0.5])
\end{verbatim}
To couple this conduction band to the impurity, we define
the tunneling amplitude,
\begin{verbatim}
    hybridization = 0.02
    tunneling = Dict(
        "F32" => ComplexF64[sqrt(2*hybridization/pi);;]
    )
\end{verbatim}
We have thus built a simple coupling model where the uranium
impurity has a constant hybridization $\Delta = 0.01 \pi D$
with a flat band of conduction electrons, which is the
default when the optional argument \texttt{channels\_dos} is
not passed to \texttt{nrgfull}.
\par
As a last step in the preparation, we need to define some
symmetry-related information and the numerical parameters:
\begin{verbatim}
    symmetry = "D"
    identityrep = "A1"
    clebschgordan_path = "clebschgordan"
    cutoff = 700
    iterations = 50
    L = 10.0
\end{verbatim}
These variables establish, respectively, the $G_\text{UJ}$
symmetry of the system, the identity irrep $A_1$ for the
double group $\tilde O$, the directory with the
Clebsch-Gordan coefficients, the multiplet cutoff, the
number of NRG iterations to perform, and the discretization
parameter $\Lambda$.
\par
Finally, we perform an NRG calculation of the impurity
contribution to thermodynamic quantities by running
\begin{verbatim}
    for calculation in ["CLEAN","IMP"]
        nrgfull(
            symmetry,
            label,
            L,
            iterations,
            cutoff,
            shell_config,
            tunneling;
            calculation=calculation,
            clebschgordan_path=clebschgordan_path,
            identityrep=identityrep,
            spectrum=spectrum,
            lehmann_iaj=lehmann_iaj,
            channels_dos=channels_dos
        )
    end
\end{verbatim}
By running \texttt{nrgfull} with \texttt{calculation} set to
\texttt{"CLEAN"} and \texttt{"IMP"} we are calculating
thermodynamic functions for the system without the impurity
and with the impurity, respectively, whereby the impurity
contribution is obtained as the difference between the two.
The procedure for thermodynamic calculations, including the
standard numerical parameters, is described in detail in
Ref. \cite{bulla2008}. All the available inputs, including
those not shown in this example, are described in the manual
provided with the code.
\par
The familiarity with the Julia language required from the
user is minimal. The required variable types are described briefly in
the manual and more thoroughly in the Julia documentation
\cite{juliadoc}. All the input described in this section can
be included in a script, which we refer to as
\texttt{nrg.jl}, to be run from the shell with \texttt{julia
nrg.jl} or from a Julia REPL with
\texttt{include("nrg.jl")}. Examples scripts for several
systems are shipped with the code.

\subsection{Results and
performance}\label{results_and_performance}
\begin{figure}
    \centering
    \includegraphics[width=\linewidth]{"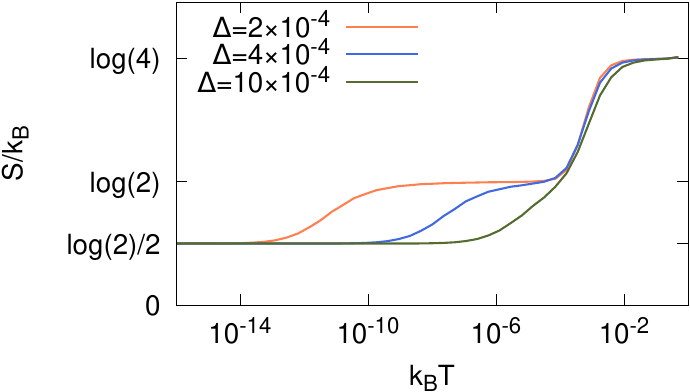"}
    \caption{Impurity contribution to the entropy $S/k_B$
    as a function of temperature for various values of
    the hybridization $\Delta=\tfrac{1}{2}\pi 
    [V(\Gamma_t)_{11}]^2$.}
    \label{fig:entropy}
\end{figure}

Using the input in Section \ref{usage} with various
hybridization values, we obtain the impurity contributions
to the entropy $S/k_B$ shown in Fig. \ref{fig:entropy}. For
the lowest hybridization value $\Delta=2\times10^{-4}$, the
system goes through all the fixed points: At high
temperatures the system behaves as a free impurity (FI) with
all the states degenerate, the entropy value $\log(4)$
indicating that both the ground and excited impurity
multiplets are thermally populated. Around $k_B
T=10\times10^{-2}$ the crossover starts from the FI regime
to the quadrupolar doublet (QD) regime, where the entropy
$\log(2)$ corresponds to a free $\Gamma_g$ ground multiplet
with a depopulated $\Gamma_e$ excited multiplet. At $k_B
T=10^{-10}$, the system undergoes a crossover to the
non-Fermi liquid (NFL) fixed point with the characteristic
entropy value $\log(2)/2$ \cite{sire1993}. For a larger
$\Delta=4\times10^{-4}$, the QD fixed point is not fully
developed and the crossover to the NFL fixed point happens
at a higher temperature. This tendency is accentuated for
the largest $\Delta=10\times10^{-4}$, which produces a
direct, higher-temperature transition from FI to NFL without
stopping at QD. 
\par
To measure the performance of the code in this new version,
we have measured the elapsed (WALL) time and maximum
resident set size (MRSS) for single NRG runs. These runs
consist on computing the thermodynamic properties of the
"3-7-8" model with the impurity included, as opposed to
\texttt{nrg.jl} where an additional calculation for the
conduction bands is included. The data was obtained using
the GNU \texttt{time} command \cite{gnutime}. To isolate the
performance of the calculation from the package loading
overhead, we have run the code using a sysimage created with
the \texttt{PackageCompiler} package. The results are
obtained for using an Intel(R) Core(TM) i7-10750H CPU @
2.60GHz processor and 16Gb of RAM.
\par
\begin{figure}
    \centering
    \includegraphics[width=0.7\linewidth]{"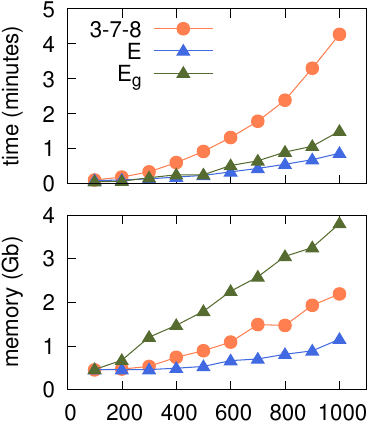"}
    \caption{
        WALL time and maximum resident set size (MRSS) for
        NRG runs with various multiplet cutoffs performed
        using the "3-7-8" model and the $E$ model with the
        new version of the code, and for the $E_g$ model
        with the previous version of the code. All
        calculations consist of 42 iterations and and use
        the same discretization parameter $\Lambda=3$.
    }
    \label{fig:performance}
\end{figure}

 
In Fig. \ref{fig:performance} we show the results obtained
for the "3-7-8" model and also for the impurity model with
cubic symmetry used in the first work
\cite{calvo-fernandez2024}, for which we compare the results
obtained with the new version for the $E$ model with the
benchmarks from the previous version (see Ref.
\cite{calvo-fernandez2024}), obtained for an equivalent
$E_g$ model. Focusing first on the $E$ and $E_g$ models, we
can see that the code is faster and more memory-efficient
than in our previous version, even when we are using more
complex data structures related to the non-simply reducible
symmetry groups. This is a direct result of the improvements
made in the algorithm, especially to those related to a
better handling of the Clebsch-Gordan summations in Eqs.
\ref{eq:D}, \ref{eq:K} and \ref{eq:F}. The scaling of both
time and memory consumption with the multiplet cutoff
remains similar to the previous version. The results for the
"3-7-8" model show that it is more time- and memory-consuming
than the $E$ model, which we attribute mainly to the
$O_\text{O}\otimes SU(2)_\text{S}$ spin-orbital symmetry of
the $E$ system, which is larger than the double-group
$\tilde O_\text{OS}$ symmetry of the "3-7-8" model and results
in a more efficient calculation \cite{calvo-fernandez2024}.

\section{Conclusions}
We have presented a new version of the \verb|PointGroupNRG|
code that allows to construct magnetic impurity models using
a standard Anderson Hamiltonian or an ionic model, and solve
it efficiently using the numerical renormalization group
method. The code fully exploits symmetries including
conservation of particles, conservation of spin,
conservation of total angular momentum and, importantly,
point and double group symmetries. Using as an example the
so-called uranium "3-7-8" model, which gives rise to the
quadrupolar Kondo effect, we showed that the code is
applicable to realistic models for systems with
crystal-field effects and spin-orbit interaction.



%% References

\section*{Acknowledgments}
We acknowledge grants No. IT-1527-22, funded by the
Department of Education, Universities and Research of the
Basque Government; PID2022-137685NB-I00, funded by MCIN/AEI
10.13039/501100011033/ and by “ERDF A way of making Europe”,
and PRE2020-092046, funded by the Spanish MCIN.  

\bibliographystyle{elsarticle-num}
\bibliography{main.bib}






% \end{thebibliography}

\end{document}